\documentclass{article}
\usepackage{spconf,amsmath,graphicx}
 \usepackage{booktabs}
 \usepackage[thinc]{esdiff}
 \usepackage{multirow,xcolor}
 \usepackage{subfig}
\newcommand{\be}{\begin{eqnarray}}
\newcommand{\ee}{\end{eqnarray}}

\usepackage{comment}
\usepackage{multirow}
\usepackage{graphicx}
\usepackage{url}
\usepackage{xcolor,colortbl}

\definecolor{Gray}{gray}{0.85}
\definecolor{LightCyan}{rgb}{0.88,1,1}

\title{End-to-end speech recognition with joint \\ dereverberation of sub-band autoregressive envelopes  
}
\name{Rohit Kumar$^\dagger$, Anurenjan Purushothaman$^{\dagger \ddagger}$, Anirudh Sreeram$^\#$, Sriram Ganapathy$^\dagger$\thanks{This work was funded by the project grants from Samsung Research India, Bangalore, India.}}

\address{
  $^\dagger$Learning and Extraction of Acoustic Patterns (LEAP) lab, Indian Institute of Science, Bangalore. \\
  $^\ddagger$College of Engineering, Thiruvanthapuram, India. \\
  $^\#$University of Southern California, Los Angels, USA.\\
  E-mail - {\{rohitk, anurenjanr,sriramg\}@iisc.ac.in, asreeram@usc.edu}
  }
\ninept 

\begin{document}

\maketitle
\begin{abstract}
The end-to-end (E2E) automatic speech recognition (ASR) systems are often required to operate  in reverberant  conditions, where the long-term sub-band envelopes of the speech are temporally smeared. 
In this paper, we develop a feature enhancement approach using a neural model  operating on sub-band  temporal envelopes. The temporal envelopes are modeled using the  framework of frequency domain linear prediction (FDLP).  The neural enhancement model proposed in this paper performs an envelope gain based enhancement of temporal envelopes. The model architecture consists of a combination of convolutional and long short term memory (LSTM) neural network layers. Further, the envelope dereverberation, feature extraction and acoustic modeling using transformer based E2E ASR can all be jointly optimized for the speech recognition task. 
%The joint optimization ensures that the dereverberation model targets the ASR cost function. 
We perform E2E speech recognition experiments on the REVERB challenge dataset as well as on the VOiCES dataset. In these experiments, the proposed joint modeling approach yields significant improvements compared to the baseline E2E ASR system (average relative improvements of $21$\% on the REVERB challenge dataset and about $10$\% on the VOiCES dataset).  \end{abstract}

%\cite{hain2012transcribing}

\section{Introduction}
In the present era of smart speakers, virtual assistants, and human-machine speech interfaces, the approach of end-to-end (E2E) automatic speech recognition (ASR) finds wide spread application due to the elegance in processing, computational simplicity and edge deployment possibilities. 
Most of these speech applications require functioning in far-field reverberant environments. The far-field recording condition smears the speech signal~\cite{ganapathy2012signal}, adversely impacting the ASR performance \cite{yu2016automatic}, with performance degradation  of up to $70$\% \cite{dan}. 

A common approach in multi-channel recording conditions is to use a weighted and delayed combination of the multiple channels using beamforming \cite{anguera2007acoustic}. The current state-of-art approaches to beamforming use a neural mask estimator~\cite{heymann2016neural,rohit}. The speech and noise mask estimations are used to derive the power spectral density of the source and interfering signals \cite{warsitz2007blind}.   Further, a weighted prediction error (WPE)~\cite{wpe} based dereverberation is used along with beamforming.   In spite of these approaches to suppress far-field effects, the temporal smearing of sub-band envelopes, causes performance degradation in ASR systems  \cite{peddinti2017low}.

%The end-to-end (E2E) speech recognition consists of a single deep neural network model which combines the acoustic and pronunciation modeling~\cite{graves2014towards}. This model is trained on words, sub-words or character  targets \cite{soltau2016neural}. The E2E models are also computationally efficient for edge device ASR applications~\cite{on_device}. 
%Given the elegant processing in E2E models, this paper attempts to propose feature enhancement models for E2E ASR systems. 

Our previous work~\cite{purushothaman2020deep, purushothaman2021csl} explored the use of dereverberation of sub-band envelopes for hybrid speech recognition systems. The sub-band envelopes are extracted using the autoregressive modeling framework of frequency domain linear prediction \cite{thomas2008recognition,ganapathy2018far}.  The deep neural enhancement model is trained to predict an envelope gain, which  is multiplied with the sub-band envelopes of the reverberant speech for dereverberation. 

In this paper, we extend the prior works for far-field E2E ASR systems, where a joint learning of enhancement model and the E2E ASR model is proposed. We explore the boundary equilibrium generative adversarial networks (BEGAN) based loss function \cite{berthelot2017began} in the envelope dereverberation model.    In various E2E ASR experiments performed on the REVERB challenge dataset \cite{reverb} as well as the VOiCES dataset \cite{voices}, we show that the proposed approach improves over the state-of-art E2E ASR systems based on log-mel features with generalized (GEV) beamforming. Further, we illustrate that the proposed approach yields the best published results on the REVERB challenge dataset. 

%weninger2015speech
\section{Literature Review}\label{sec:prior_work}
The neural approaches to speech enhancement have witnessed considerable advances in the last decade \cite{wollmer2013feature}. In early efforts, Maas et. al \cite{maas2013recurrent} proposed a recurrent model to map the  noisy features to the clean features. Santos et. al proposed a context aware recurrent neural network ~\cite{santos2018speech} for enhancing speech spectrogram.    The mapping in the time domain signal is investigated in Pandey et. al \cite{pandey_2019}, where the loss is computed in the frequency domain loss.  Recently, end-to-end models with attention based modeling have also been explored on the REVERB challenge dataset~\cite{subramanian2019investigation,zhang2020end}. 
Previously, we had proposed a convolutional neural network model to perform dereverberation of speech~\cite{purushothaman2020deep,purushothaman2021csl}.  In the current work, we extend this prior work for E2E transformer based ASR system. 

\section{Proposed Approach}\label{sec:Proposed_Approach}
\subsection{Signal model}\label{sec:signal_model}
The speech data recorded by a far-field microphone is modeled as, 
\be
\label{eq:reverb_sig}
r(n) = x(n)*h(n),
\ee
where $x(n)$, $h(n)$ and $r(n)$  are the source speech signal, the room impulse response function and the far-field speech respectively. The room response function can be further expanded as $h(n) = h_e(n) + h_l(n)$, where $h_e(n)$ and $h_l(n)$ represent the early and late reflection components. 

Let $x_q(n)$, $h_q(n)$ and $r_q(n)$ denote the sub-band clean speech, room-response and the reverberant speech respectively for the $q^{th}$ sub-band. The sub-band envelopes of far-field speech $m_{rq}(n)$, extracted using frequency domain linear prediction (FDLP) \cite{ganapathy,ganapathy2012signal}, can be approximated as, \cite{purushothaman2020deep}
\be
m_{rq}(n) \approx \frac{1}{2} m_{xq}(n) * m_{hq}(n)
\ee
where, $m_{xq}(n)$, $m_{hq}(n)$ denote the sub-band envelope of the clean source signal and the room impulse response function respectively.  Given this envelope convolution model, we can further split the far-field speech envelope into early and late reflection components. 
\be
\label{eq:envelope_conv_model_early}
m_{rq}(n) = m_{rqe}(n) + m_{rql}(n)
\ee
where $m_{rqe}(n)$ and $m_{rql}(n)$ denote the sub-band envelopes of early and late reflection parts.
%We propose the use of a neural model for suppressing the late reverberation components in the sub-band envelopes.  
 
\subsection{Envelope Dereverberation and E2E ASR}\label{sec:neural_network} 

\begin{figure}[t!]
  \centering
  \includegraphics[height=11cm, width=8cm]{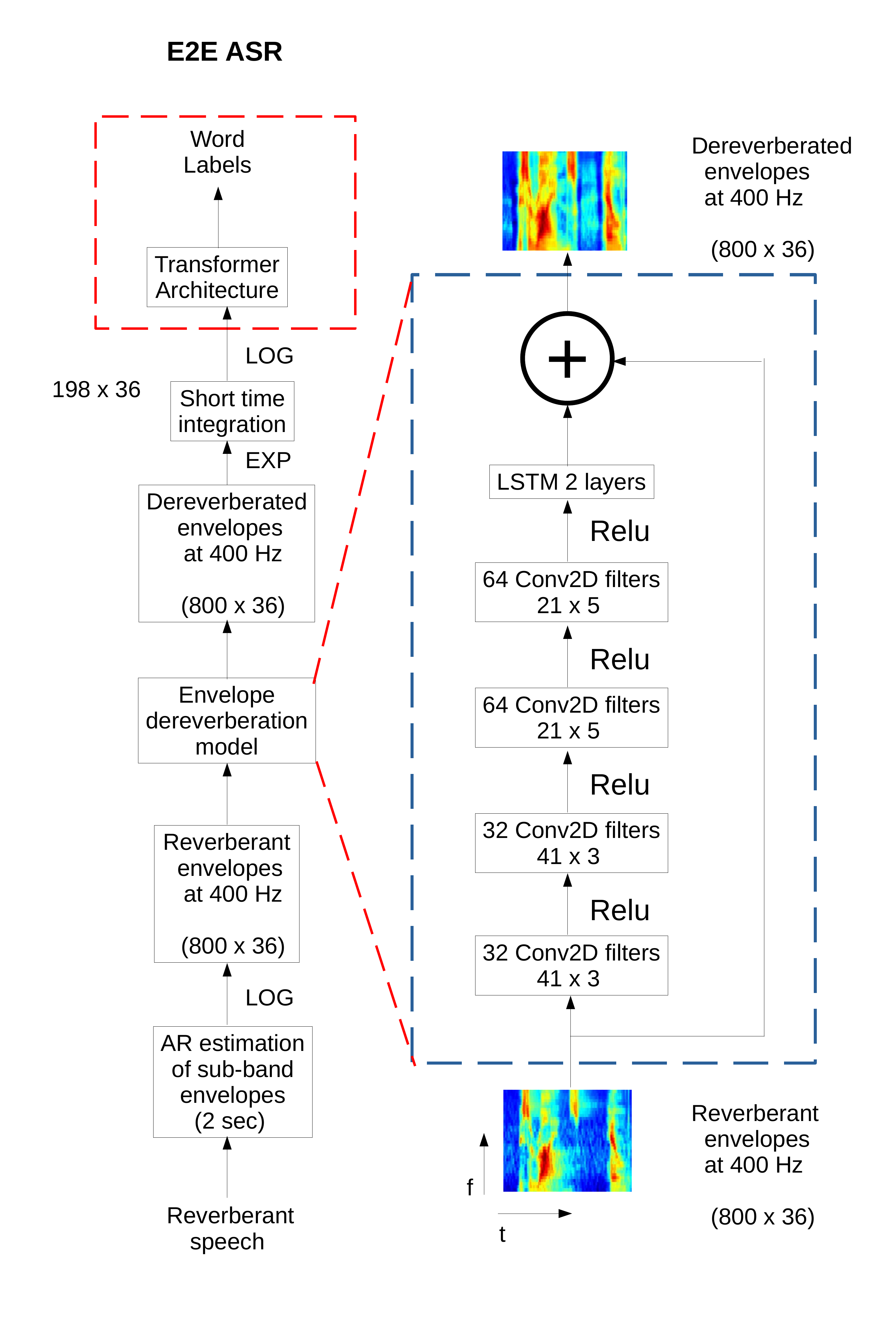}
  \vspace{-0.4cm}
      \caption{Block schematic of envelope dereverberation model, the feature extraction module and the E2E ASR model. }
  \label{fig1}
  \vspace{-0.5cm}
\end{figure}

In this section, we describe the proposed approach to envelope dereverberation (Figure \ref{fig1}).
\subsubsection{Envelope dereverberation model}
As seen in Eq.~(\ref{eq:envelope_conv_model_early}), the FDLP envelope of reverberant speech can be written as the sum of the direct component (early reflection) and those with the late reflection. The envelope dereverberation model tries to subtract the late reflection components $m_{rql}(n)$ from reverberant sub-band temporal envelope $m_{rq}(n)$. Similar to the popular Wiener filtering technique for dealing with additive noise in speech \cite{martin2005speech}, the dereverberation task is posed as an envelope gain estimation problem. The sub-band envelope residual (targets for the neural model)  is the log-difference of the sub-band envelope for the direct components and the sub-band envelope of the reverberant sub-band signal. 
The neural model is trained with reverberant sub-band envelopes ($log~(m_{rq}(n)$) as input and model outputs the gain (in the log domain this is $log~\frac {m_{rq}(n)}{m_{xq}(n)}$), which when multiplied  with the reverberant envelopes (additive in log domain), generates the estimate of source signal envelope ($log~(\hat{m}_{xq}(n)$). 
 
Figure~\ref{fig1} shows the block schematic of the proposed envelope dereverberation model.  For training the dereverberation model, we use the FDLP envelope of the close talking microphone to compute the target residual envelope. Thus, the model behaves like a residual network for dereverberation. %As the envelopes and the gain parameters are positive in nature, the model implementation in the neural architecture uses a logarithmic transform at the input and the estimated gain is followed by an exponential operation. This implementation in the log envelope domain makes the model behave like a residual network based dereverberation architecture. 

The model developed in Section~\ref{sec:signal_model} is applicable only for envelopes extracted from long analysis windows (greater than the T60 of room impulse response). Further,  the neural model also predicts the envelope residual of all sub-bands jointly to effectively utilize the sub-band correlations that exist in speech. For training the neural dereverberation model, the FDLP sub-band envelopes, corresponding to $2$ sec. non-overlapping segments of both the reverberant speech and clean speech are extracted. With a   $2$ sec. segment of FDLP envelopes, sampled at $400$ Hz, and a mel decomposition of $36$ bands, the input representation to the neural dereverberation model is of dimension $800 \times 36$.

%\input{fig}

%From the reverberant speech and the corresponding clean speech, the FDLP sub-band envelopes corresponding to $2$sec. non-overlapping segments are extracted. The choice of non-overlapping $2$sec. is due to computational considerations and the T60 values encountered in practice. If the input sampling rate is $16$ kHz, a $2$sec. segment will correspond to 32,000 samples. FDLP envelopes are extracted at a down sampled rate of $400$ Hz. Thus every $2$sec. segment of audio corresponds to $800$ samples of FDLP envelope for each sub-band. We use a $36$ band mel decomposition. This makes the representation at the input of the enhancement model of size $800 \times 36$. 
The target signal for the dereverberation model in Figure~\ref{fig1} is the log-difference between the envelope of the close talking (clean) FDLP envelopes and those of the reverberant speech.  
The final architecture of the neural model is based on convolutional long short term memory (CLSTM) networks (Figure~\ref{fig1}). The input $2$-D data of sub-band envelopes are fed to a set of convolutional layers (two layers having $32$ filters each with kernels of size   $41 \times 5$ followed by two layers of $64$ filters with size $21 \times 3$). The CNN layers do not have pooling and include zero padding to preserve the input size.  The output of the CNN layers are input to $3$ layers of LSTM cells with $1024,1024$, and $36$ units respectively. The last layer size is matched with that of the target signal (log-difference). The training criteria is based on the mean square error between the target and predicted model output. The model is trained with stochastic gradient descent using Adam optimizer. 
% Please add the following required packages to your document preamble:
% \usepackage{multirow}
% \usepackage{graphicx}
\begin{table*}[]
\centering
\caption{WER (\%) in the REVERB dataset for different  dereverberation model  architectures  with  BF-FDLP features. All the features are used with transformer based E2E acoustic model with separate training of the dereverberation module and  E2E ASR module.}
\vspace{-0.2cm}
\label{tab:my-table}
\resizebox{0.85\textwidth}{!}{%
\begin{tabular}{c|c|c|c|c|c|c|c|c}
\hline
\multirow{3}{*}{\textbf{\begin{tabular}[c]{@{}c@{}}Model\\ Architecture\end{tabular}}} & \multirow{3}{*}{\textbf{\begin{tabular}[c]{@{}c@{}}Derevb. Model\\ Parameters\\ (in Million)\end{tabular}}} & \multirow{3}{*}{\textbf{\begin{tabular}[c]{@{}c@{}}Loss\\ Function\end{tabular}}} & \multicolumn{3}{c|}{\textbf{DEV}} & \multicolumn{3}{c}{\textbf{EVAL}} \\ \cline{4-9} 
 &  &  & \multirow{2}{*}{\textbf{Real}} & \multirow{2}{*}{\textbf{Sim}} & \multirow{2}{*}{\textbf{Avg}} & \multirow{2}{*}{\textbf{Real}} & \multirow{2}{*}{\textbf{Sim}} & \multirow{2}{*}{\textbf{Avg}} \\
 &  &  &  &  &  &  &  &  \\ \hline
 
%\rowcolor{Gray}
2 CNN + Transformer & 54.7 & MSE & 15.1 & 10.7 & 12.9 & 12.2 & 9.9 & 11.1 \\ %\hline

4 CNN + Transformer & 137.4 & MSE & 14.8 & 9.8 & 12.3 & 11.3 & 9.7 & 10.5 \\ %\hline

% \rowcolor{Gray}
Linear Layer + Transformer & 19.8 & MSE & 14.3 & 9.5 & 11.9 & 11.5 & 9.3 & 10.4 \\ 

4 CNN + 2 LSTM &  14.5 & MSE &  11.4 &  7.7 &  9.5 &  9.5 &  7.0 &  8.2 \\  \midrule 
4 CNN + 2 LSTM  & 14.5 & MSE + 0.2*BEGAN & \textbf{11.2} & 8.2 & 9.7 & 12.5 & 9.3 & \multicolumn{1}{l}{10.9} \\ %\hline
4 CNN + 2 LSTM & 14.5 & MSE + 0.1*BEGAN & 11.3 & \textbf{7.5} & \textbf{9.4} & \textbf{8.7} & \textbf{6.6} & \multicolumn{1}{l}{\textbf{7.6}} \\ \hline
\end{tabular}%
}
\vspace{-0.2cm}
\label{table:3}
\end{table*} 
We also experiment with the learning of the model with boundary equilibrium generative adversarial network (BEGAN) loss~\cite{berthelot2017began}. 
The BEGAN ~\cite{berthelot2017began} is an energy based GAN. The model attempts to match the distribution of loss function using an auto-encoder (AE) architecture. The model uses the equilibrium  of AE loss using the hyper-parameter $\gamma \in [0,1]$, termed as the diversity ratio.  
We use the BEGAN discriminator to introduce the adversarial setting in the training process. 
%Our experiments   show that BEGAN improves the quality of the reconstruction output when used in the setting with separate dereverberation and E2E ASR modules. However, the experiments with joint training of the dereverberation network and the E2E ASR network did not show performance improvements for training with the BEGAN based regularization. 
%\subsubsection{Feature Extraction and Acoustic Modeling}

The predicted sub-band envelope from the dereverberation model ($\hat{m}_{xq}(n)$) for each band is integrated in short Hamming shaped windows of size $25$ ms with a shift of $10$ ms \cite{ganapathy2012signal}.   A log compression is applied to limit the dynamic range of values. The integrated sub-band envelopes are input to the ASR as 2-D time frequency features of the audio signal. It is noteworthy that the feature extraction steps (Hamming window based integration and log compression) can be represented as fixed neural layers consisting of  a $1$-D CNN layer.  Hence, the entire set of steps, starting from FDLP envelope extraction to the E2E ASR transcript generation, can be realized using differentiable neural operations\footnote{\textcolor{black}{The implementation of the work can be found in \url{https://github.com/iiscleap/Joint_FDLP_envelope_dereverberation_E2E_ASR/tree/master}}}. 
\begin{table}[t!]
\caption{Word Error Rate (\%) in REVERB dataset for different end-to-end architectures.}
\vspace{-0.5cm}
\begin{center}
\resizebox{0.8\columnwidth}{!}{%
\begin{tabular}{@{}l|c|c|c@{}}
\toprule
\multicolumn{1}{c|}{\textbf{\begin{tabular}[c]{@{}c@{}}Model\\ features\end{tabular}}} & \textbf{\begin{tabular}[c]{@{}c@{}}Model\\ architecture\end{tabular}} & \textbf{Dev avg} & \textbf{Eval avg} \\ \midrule
BF-FBANK & VGG (E2E) & 14.9 & 11.7 \\
BF-FDLP & VGG (E2E) & 12.1  & 9.7  \\ \hline
BF-FBANK & Transformer (E2E) & 12.9 & 10.4 \\
BF-FDLP & Transformer (E2E) & 10.4  & 8.3  \\
\bottomrule
% BF-FBANK & CLSTM (Hybrid) \cite{anu1} & 9.4 & 10.6 \\
% BF-FDLP & CLSTM (Hybrid) \cite{anu1} & 12.3  & 10.5 \\
% \bottomrule
\end{tabular}}
\end{center}
\label{table:2}
\vspace{-0.5cm}
\end{table}

\subsubsection{E2E ASR framework}

We use the ESPnet toolkit \cite{watanabe2018espnet} to perform all the end-to-end speech recognition experiments, with the Pytorch backend~\cite{pytorch}. We experimented with two end to end model architectures, (i) VGG based model and (ii) transformer architecture \cite{karita2019comparative}. The first E2E model architecture uses $3$-layer VGG-BLSTM based encoder with $1024$ units, and $1$-layer of decoder with $1024$ units. In the transformer architecture, the encoder used is a $12$-layer transformer network with $2048$ units in the projection layer. The encoder-decoder attention is used where the decoder network is a $6$-layer transformer architecture with $2048$ units in the  projection layer. During training, a combination of multiple cost functions is used \cite{karita2019comparative} which consists of connectionist temporal cost (CTC) loss and the  attention based cross entropy (CE) loss. The CTC-weight is fixed at $0.3$ and during decoding the beam-size is fixed at $10$. The model is trained  for several epochs till the loss function saturates on the validation data, with the patience factor of $2$ epochs for REVERB dataset and $3$ epochs for VOiCES dataset. 
%In order to alleviate over fitting, label smoothing techniques are applied while training. 

A recurrent neural network based language model (RNN-LM) with $1$ layer of $1000$ LSTM cells is employed. The stochastic gradient descent (SGD) optimizer with a batch of $32$ is used to train the model. The language model is incorporated in the end-to-end system and we have augmented the training data with clean Wall Street Journal (WSJ)  data. 
%However, no data augmentation is used for training the envelope dereverberation model. 

% In our experiments with reverb dataset, we observed that the second E2E transformer based model architecture was performing better than the VGG-BLSTM based model, and hence performed further experiments with the second model.

%\footnote{We can add these results if they are available with us}.

%\input{Tables/Table_2_rev_diff_arch}

\subsubsection{Joint learning}
The joint learning of the  envelope dereverberation module and the E2E ASR architecture is achieved by constructing the single neural model (as shown in Figure~\ref{fig1}). Given the deep structure consisting of convolutions, LSTMs and transformer based layers, we initialize the modules with isolated training of each component. Specifically, the envelope dereverberation model is trained using MSE+BEGAN loss and the E2E architecture is  separately trained on the acoustic features from the envelope dereverberation model. 
%The pre-trained envelope dereverberation module and E2E architecture based ASR model are combined with the neural feature extraction layers. 
The final model is jointly optimized using  the E2E ASR loss function (combined CTC and CE loss). The audio signal is divided into non overlapping segments of $2$ sec.  length and  passed through the envelope dereverberation model. 
%After applying short time integration on those $2$ sec chunks, we apply global mean variance normalization (MVN).
The  feature vectors for the $2$ sec audio chunk are passed through the E2E  architecture to predict the acoustic model targets.

\section{Experiments and results}\label{sec:expt}
%The experiments are performed on REVERB challenge~\cite{reverb} and VOiCES datasets~\cite{voices_inter}. 
For all the models, we use WPE enhancement~\cite{wpe} along with unsupervised generalized eigen value (GEV) beamforming~\cite{warsitz2007blind}. The baseline features are the filter-bank energy features (denoted as BF-FBANK). Since VOiCES is a single channel dataset, we use only the WPE enhancement.   The FBANK features are $36$ band log-mel spectrogram with frequency range from $200$ Hz to $6500$ Hz (similar to the sub-band decomposition in the FDLP feature extraction). 

\subsection{REVERB Challenge ASR}
%\subsubsection{Data}
The REVERB challenge dataset \cite{reverb} for ASR consists of $8$ channel recordings with real and simulated reverberation conditions. The  simulated data is comprised of reverberant utterances (from the WSJCAM0 corpus \cite{rev1}) obtained by artificially convolving clean WSJCAM0 recordings with the measured room impulse responses (RIRs) and adding noise at an SNR of $20$ dB. The simulated data has six different reverberation conditions. The real data, which is comprised of utterances from the MC-WSJ-AV corpus \cite{rev2}, consists of utterances spoken by human speakers in a noisy reverberant room. The training set consists of $7861$ utterances from the clean WSJCAM0 training data convolved with $24$ measured RIRs. 
%More details about the dataset are available in \cite{rev3}. 

%\input{Tables/Table_2_rev_diff_arch}

%\input{Tables/Table_6}
%\input{Tables/table_with_param_rev}

%\subsubsection{Results and discussion}
The first experiments report the performance of the two E2E architectures explored in this paper. We experiment with baseline features (BF-FBANK) and the FDLP features without dereverberation for these experiments. These results are reported in Table~\ref{table:2}. The transformer based E2E architecture shows significant improvements over the VGG based model. The rest of the experiments reported in the paper use the transformer architecture for the E2E model training. Further, the BF-FDLP results are observed to perform consistently better than the BF-FBANK baseline (average absolute improvements of $2.5$\% on the development set and about $2.1$\% on the evaluation set). These improvements may be attributed to the advantages of autoregressive modeling of sub-band envelopes, where the signal peaks are given more prominence \cite{ganapathy2012signal}. 

\begin{table}[t!]
\caption{WER (\%) in REVERB dataset for separate learning of the dereverberation and E2E models as well as the joint learning.}
\vspace{-0.4cm}
\begin{center}
{
\centering
\resizebox{0.98\columnwidth}{!}{%
\begin{tabular}{@{}l|ccc|ccc@{}}
\toprule
\multicolumn{1}{l|}{\multirow{2}{*}{\textbf{\begin{tabular}[c]{@{}c@{}}
Model\\ Config.
\end{tabular}}}} &   \multicolumn{3}{c|}{\textbf{Dev}}                                              & \multicolumn{3}{c}{\textbf{Eval}}                                             \\ \cmidrule(l){2-7} 
\multicolumn{1}{c|}{}                                                                                   &  \multicolumn{1}{l}{\textbf{Real}} & \multicolumn{1}{l}{\textbf{Sim}} & \multicolumn{1}{l|}{\textbf{Avg}} & \multicolumn{1}{l}{\textbf{Real}} & \multicolumn{1}{l}{\textbf{Sim}} & \multicolumn{1}{l}{\textbf{Avg}} \\ \midrule
BF-FBANK  (baseline)  &  15.3                     & 10.5                      & 12.9 & 11.5                     & 9.2                      & 10.4                    \\
BF-FDLP  &     14.1                     & 6.7                      & 10.4                     & 10.1                     & 6.5                      & 8.3                    \\
- + derevb. [MSE] &    {11.4 }                    & {7.7 }                     & {9.5}                      & {9.5}                     & 7.0                      & {8.2} \\ 
- +  derevb. [MSE+BEGAN] & 11.3 & 7.5 & {9.4}                      & {8.7}                     & 6.6                      & {7.6} \\ \midrule
- +  joint. [MSE]   & {10.3}                    & {6.3}                     & {8.3}                      & \textbf{7.1}                     &\textbf{5.6}                       & \textbf{6.3}                      \\  
- +  joint. [MSE+BEGAN] &                                                                                             {9.3}                    & {6.1}                     & \textbf{7.7}                      & {7.7}                     &{5.9}                       & {6.8} 
\\ \bottomrule

\end{tabular}}}

\label{table:1}
\end{center}
\vspace{-0.4cm}
\end{table}
The next set of experiments report the performance for various envelope dereverberation model architectures. In these experiments, we perform a separate dereverberation and speech recognition E2E model training. Table \ref{table:3} shows the results for the different models that are used to perform the envelope dereverberation. The $2$ CNN+transformer model employs two CNN layers with $32$ filters each with the kernel size of $41\times5$, and four layers of  transformer encoder architecture with $8$ attention heads. The $4$ CNN+transformer model, uses the same transformer architecture, and the same initial two layers of CNN with the only difference being two additional CNN layers. Here, the $2$ additional layers of CNN contain $64$ filters each and with kernel size of $21\times3$. In the model defined as  linear+transformer, we use the same transformer configuration, but use a simple linear layer to project the feature matrix, which is passed through the transformer. Further, the architecture with $4$ CNN and $2$ LSTM layers gave the best performance (similar to the previous findings on hybrid ASR model~\cite{purushothaman2020deep}). %Also, as seen in Table~\ref{table:3}, when the number of trainable model parameters were reduced, the performance of the model was observed to improve. This may also indicate that the transformer based architectures for dereverberation network may suffer from over-fitting. 

The last two rows of Table~\ref{table:3} highlight the performance with regularized loss function (MSE + BEGAN loss). As seen here, the regularization with $0.1$ parameter for BEGAN loss results in the best performance reported in this Table. In particular, on the real evaluation data, the model trained with MSE and BEGAN loss improves over the MSE alone training. We observe a $27$\% relative improvement in the development and evaluation dataset compared to the baseline BF-FBANK  (Table~\ref{table:2}).

The results using the joint learning of the dereverberation network and the E2E ASR model are reported in Table \ref{table:1}.  The joint training model is initialized using the deverberation model and the E2E model trained separately. In these experiments, we use the deverberation network with $4$ CNN layers and $2$ LSTM layers, termed as CLSTM network.  The proposed joint training of the model yields average absolute improvements of $4.6$\% and $4.1$\% on the development set and evaluation set respectively over the baseline system. The improvement in real condition is  more than those observed in the simulated data. The joint model, initialized with the dereverberation model trained with BEGAN loss regularization, improved over the one without the BEGAN loss regularization, in the development data. However, this did not show consistent improvement in the evaluation data.

For the joint model initialized with the MSE loss based CLSTM dereverberation network, we observe average relative improvements of $36$\% on the development set and about $39$\% on the evaluation set, compared to the BF-FBANK baseline. The joint training is also shown to improve over the set up of having separate networks for dereverberation and E2E ASR.  These results show that the joint learning of the two modules and the application of autoregressive modelling of sub-band envelopes can yield considerable benefits for E2E ASR. The comparison of the results from prior works reported on the REVERB challenge dataset is given in Table~\ref{table:reverb_PriorWorks}. The table includes results from end-to-end ASR systems~\cite{subramanian2019investigation,zhang2020end,fujita2020attention} as well as the joint enhancement and ASR modeling work reported in \cite{heymann2019joint}. To the best of our knowledge, the results from the proposed algorithm achieves the best average performance on the REVERB challenge evaluation dataset (relative improvements of $15$\% over the recent work by Fujita et. al.~\cite{fujita2020attention}). 

\begin{table}[t!]
\centering
\caption{Comparison of the results with other works reported on REVERB challenge dataset. }
\vspace{-0.2cm}
\resizebox{0.8\columnwidth}{!}{%
\begin{tabular}{l|c|c|c}
\toprule
System & Eval-sim. & Eval-real & Avg. \\   \hline       
Subramanian et. al.~\cite{subramanian2019investigation} & 6.6 & 10.6 & 8.6 \\  
Heymann et. al.~\cite{heymann2019joint}      & -              & 10.8  & - \\    
Fujita et. al.~\cite{fujita2020attention} & \textbf{4.9 }& 9.8 & 7.4\\ 
Zhang et. al.~\cite{zhang2020end}      & -              & 10.0  & - \\  \hline 
This work & 5.6  & \textbf{7.1} &\textbf{ 6.3} \\
\bottomrule
\end{tabular}}
\vspace{-0.2cm}
 \label{table:reverb_PriorWorks}
\end{table}

\subsection{VOiCES ASR}
%\subsubsection{Data}
The training set of the VOiCES corpus \cite{voices} consists of $80$-hour subset of the clean LibriSpeech corpus. The training set has close talking microphone recordings from $427$ speakers recorded in clean environments. The development and evaluation set consists of $19$ hours and $20$ hours of far-field microphone recordings from diverse room dimensions, environment and noise conditions. There are no common speakers between the training set and the development set or the  evaluation set. The significant  mismatch the training set and development/evaluation set allows the testing of the robustness of the trained models. We have used the same transformer based E2E ASR system that was developed for the REVERB challenge dataset. Further, the current experiments do not perform any data augmentation in the ASR model training. 

%\subsubsection{Results and discussion}
\begin{table}[t!]
\centering
\caption{Performance (WER \%) on the VOiCES dataset.}
\vspace{-0.2cm}
\resizebox{0.55\columnwidth}{!}{%
\begin{tabular}{l|c|c}
\toprule
\multirow{1}{*}{\textbf{Model Config.}} & \multicolumn{1}{c|}{\textbf{Dev}}            & \multicolumn{1}{c}{\textbf{Eval}}            
                                      \\ \midrule
FBANK                             &42.9              &52.5          \\
FDLP                            &40.2            & 49.9        \\
FDLP + CLSTM derevb.                             &39.2            & 48.6        \\ \hline
~~~~~~~~~~~~~ + joint. (prop)                           & \textbf{38.1}           & \textbf{47.6}        \\
\bottomrule
\end{tabular}}
\vspace{-0.5cm}
\label{table:voice_table}
\end{table}

 The WER results for VOiCES corpus is given in Table \ref{table:voice_table}. As seen, the FDLP features show  better WER compared to the FBANK features. This is improved with the dereverberation of the FDLP envelopes, achieved using the CLSTM network. The same architecture of the dereverbation network used in the REVERB dataset is also explored for the VOiCES dataset. In this case, the dereverberation network was trained with simulated reverberation. 
  
 The best results reported in Table~\ref{table:voice_table} is for the model with joint learning of the  dereverberation network and the E2E model. For training the joint E2E model, we initialize the weights of encoder and decoder with weights of CLSTM E2E model, and then train the joint model for $5$ epochs. The final WER shows an average relative WER improvement of $11.2$\% in  development set and $9.3$\% on the evaluation set over the baseline FBANK system.

\section{Summary}\label{sec:summary}
In this paper, we propose a feature enhancement model for E2E ASR systems using frequency domain linear prediction based sub-band envelopes. Using the joint learning of the neural dereverberation approach and the E2E ASR model, we perform several speech recognition experiments on the  REVERB challenge dataset as well as on the VOiCES dataset. These results shows that the proposed approach improves over the state of art E2E ASR systems based on log mel features. Further, ablation studies show the justification for the choice of the dereverberation network architecture and the choice of loss functions in training the models. 

%In this paper attempts to propose feature enhancement models for E2E ASR systems. Using the neural enhancement approach, we perform E2E speech recognition experiments on the  REVERB challenge dataset as well as on the VOiCES dataset, and showed that the proposed approach improves over the state of art E2E ASR systems based on log mel features with GEV beamforming and WPE enhancement

\ninept
\bibliographystyle{IEEEbib}
\bibliography{mybib}

\begin{thebibliography}{10}

\bibitem{ganapathy2012signal}
S.~Ganapathy,
\newblock {\em Signal analysis using autoregressive models of amplitude
  modulation},
\newblock Ph.D. thesis, Johns Hopkins University, 2012.

\bibitem{yu2016automatic}
D.~Yu and Li~Deng,
\newblock {\em Automatic Speech Recognition.},
\newblock Springer, 2016.

\bibitem{dan}
V.~{Peddinti}, Y.~{Wang}, D.~{Povey}, and S.~{Khudanpur},
\newblock ``Low latency acoustic modeling using temporal convolution and
  {LSTM}s,''
\newblock {\em IEEE Signal Processing Letters}, vol. 25, no. 3, pp. 373--377,
  2018.

\bibitem{anguera2007acoustic}
X.~Anguera, C.~Wooters, and J.~Hernando,
\newblock ``Acoustic beamforming for speaker diarization of meetings,''
\newblock {\em IEEE TASLP}, vol. 15, no. 7, pp. 2011--2022, 2007.

\bibitem{heymann2016neural}
Jahn Heymann, Lukas Drude, and Reinhold Haeb-Umbach,
\newblock ``Neural network based spectral mask estimation for acoustic
  beamforming,''
\newblock in {\em IEEE International Conference on Acoustics, Speech and Signal
  Processing (ICASSP)}, 2016, pp. 196--200.

\bibitem{rohit}
R.~{Kumar}, A.~{Sreeram}, A.~{Purushothaman}, and S.~{Ganapathy},
\newblock ``Unsupervised neural mask estimator for generalized eigen-value
  beamforming based {ASR},''
\newblock in {\em IEEE ICASSP}, 2020, pp. 7494--7498.

\bibitem{warsitz2007blind}
E.~Warsitz and R.~Haeb-Umbach,
\newblock ``Blind acoustic beamforming based on generalized eigenvalue
  decomposition,''
\newblock {\em IEEE Transactions on audio, speech, and language processing},
  vol. 15, no. 5, pp. 1529--1539, 2007.

\bibitem{wpe}
Tomohiro Nakatani, Takuya Yoshioka, Keisuke Kinoshita, Masato Miyoshi, and
  Biing-Hwang Juang,
\newblock ``Speech dereverberation based on variance-normalized delayed linear
  prediction,''
\newblock {\em IEEE TASLP}, vol. 18, no. 7, pp. 1717--1731, 2010.

\bibitem{peddinti2017low}
V.~Peddinti, Y.~Wang, D.~Povey, and S.~Khudanpur,
\newblock ``Low latency acoustic modeling using temporal convolution and
  {LSTM}s,''
\newblock {\em IEEE Signal Processing Letters, vol. 25, issue 3, pp. 373-377},
  2017.

\bibitem{purushothaman2020deep}
Anurenjan Purushothaman, Anirudh Sreeram, Rohit Kumar, and Sriram Ganapathy,
\newblock ``{Deep Learning Based Dereverberation of Temporal Envelopes for
  Robust Speech Recognition},''
\newblock in {\em Proc. Interspeech 2020}, 2020, pp. 1688--1692.

\bibitem{purushothaman2021csl}
Anurenjan Purushothaman, Anirudh Sreeram, Rohit Kumar, and Sriram Ganapathy,
\newblock ``Dereverberation of autoregressive envelopes for far-field speech
  recognition,''
\newblock {\em Computer Speech \& Language}, vol. 72, pp. 101277, 2022.

\bibitem{thomas2008recognition}
S.~Thomas, S.~Ganapathy, and H.~Hermansky,
\newblock ``Recognition of reverberant speech using frequency domain linear
  prediction,''
\newblock {\em IEEE Signal Processing Letters}, vol. 15, pp. 681--684, 2008.

\bibitem{ganapathy2018far}
S.~Ganapathy and M.~Harish,
\newblock ``Far-field speech recognition using multivariate autoregressive
  models.,''
\newblock in {\em Interspeech}, 2018, pp. 3023--3027.

\bibitem{berthelot2017began}
D.~Berthelot, T.~Schumm, and L.~Metz,
\newblock ``{BEGAN}: Boundary equilibrium generative adversarial networks,''
\newblock {\em arXiv preprint arXiv:1703.10717}, 2017.

\bibitem{reverb}
Keisuke Kinoshita et~al.,
\newblock ``The reverb challenge: A common evaluation framework for
  dereverberation and recognition of reverberant speech,''
\newblock in {\em IEEE WASPAA}, 2013, pp. 1--4.

\bibitem{voices}
C.~Richey, M.~Barrios, et~al.,
\newblock ``{VOiCES} obscured in complex environmental settings (voices)
  corpus,''
\newblock {\em arXiv preprint arXiv:1804.05053}, 2018.

\bibitem{wollmer2013feature}
M.~W{\"o}llmer, Z.~Zhang, F.~Weninger, B.~Schuller, and G.~Rigoll,
\newblock ``Feature enhancement by bidirectional lstm networks for
  conversational speech recognition in highly non-stationary noise,''
\newblock in {\em IEEE ICASSP}, 2013, pp. 6822--6826.

\bibitem{maas2013recurrent}
A.~Maas, T.~O’Neil, A.~Hannun, and A.~Ng,
\newblock ``Recurrent neural network feature enhancement: The 2nd chime
  challenge,''
\newblock in {\em Proceedings The 2nd CHiME Workshop on Machine Listening in
  Multisource Environments held in conjunction with ICASSP}, 2013, pp. 79--80.

\bibitem{santos2018speech}
J.~Santos and T.~Falk,
\newblock ``Speech dereverberation with context-aware recurrent neural
  networks,''
\newblock {\em IEEE/ACM Transactions on Audio, Speech, and Language
  Processing}, vol. 26, no. 7, pp. 1236--1246, 2018.

\bibitem{pandey_2019}
A.~{Pandey} and D.~{Wang},
\newblock ``A new framework for {CNN}-based speech enhancement in the time
  domain,''
\newblock {\em IEEE/ACM Transactions on Audio, Speech, and Language
  Processing}, vol. 27, no. 7, pp. 1179--1188, 2019.

\bibitem{subramanian2019investigation}
A.~Subramanian, X.~Wang, S.~Watanabe, T.~Taniguchi, D.~Tran, and Y.~Fujita,
\newblock ``An investigation of end-to-end multichannel speech recognition for
  reverberant and mismatch conditions,''
\newblock {\em arXiv preprint arXiv:1904.09049}, 2019.

\bibitem{zhang2020end}
W.~Zhang, A.~Subramanian, X.~Chang, S.~Watanabe, and Y.~Qian,
\newblock ``End-to-end far-field speech recognition with unified
  dereverberation and beamforming,''
\newblock {\em arXiv preprint arXiv:2005.10479}, 2020.

\bibitem{ganapathy}
S.~Ganapathy and V.~Peddinti,
\newblock ``3-d {CNN} models for far-field multi-channel speech recognition,''
\newblock in {\em IEEE ICASSP}, 2018, pp. 5499--5503.

\bibitem{martin2005speech}
R.~Martin,
\newblock ``Speech enhancement based on minimum mean-square error estimation
  and supergaussian priors,''
\newblock {\em IEEE transactions on speech and audio processing}, vol. 13, no.
  5, pp. 845--856, 2005.

\bibitem{watanabe2018espnet}
S.~Watanabe, T.~Hori, et~al.,
\newblock ``Espnet: End-to-end speech processing toolkit,''
\newblock {\em arXiv preprint arXiv:1804.00015}, 2018.

\bibitem{pytorch}
A.~Paszke, S.~Gross, et~al.,
\newblock ``Automatic differentiation in pytorch,''
\newblock in {\em NIPS-W}, 2017.

\bibitem{karita2019comparative}
S.~Karita, N.~Chen, et~al.,
\newblock ``A comparative study on transformer vs {RNN} in speech
  applications,''
\newblock in {\em 2019 IEEE ASRU}. IEEE, 2019, pp. 449--456.

\bibitem{rev1}
T.~Robinson, J.~Fransen, D.~Pye, J.~Foote, and S.~Renals,
\newblock ``{WSJCAMO}: {A} {B}ritish english speech corpus for large vocabulary
  continuous speech recognition,''
\newblock in {\em IEEE ICASSP}, 1995, vol.~1, pp. 81--84.

\bibitem{rev2}
M.~Lincoln, I.~McCowan, J.~Vepa, and H.~Maganti,
\newblock ``The multi-channel wall street journal audio visual corpus
  ({MC-WSJ-AV}): Specification and initial experiments,''
\newblock in {\em IEEE ASRU}, 2005, pp. 357--362.

\bibitem{fujita2020attention}
Y.~Fujita, A.~Subramanian, M.~Omachi, and S.~Watanabe,
\newblock ``Attention-based asr with lightweight and dynamic convolutions,''
\newblock in {\em ICASSP}. IEEE, 2020, pp. 7034--7038.

\bibitem{heymann2019joint}
J.~Heymann, L.~Drude, R.~Haeb-Umbach, K.~Kinoshita, and T.~Nakatani,
\newblock ``Joint optimization of neural network-based wpe dereverberation and
  acoustic model for robust online {ASR},''
\newblock in {\em ICASSP}. IEEE, 2019, pp. 6655--6659.

\end{thebibliography}
\end{document}